\documentclass[
    reprint,
    amsmath,
    amssymb,
    aps,
    prl,
    floatfix,
]{revtex4-2}

\usepackage{silence}
\usepackage{graphicx}
\usepackage{dcolumn}
\usepackage{bm}
\usepackage{color}

\usepackage[
    colorlinks=true,
    linkcolor=blue,
    citecolor=blue,
    urlcolor=blue
]{hyperref}

\begin{document}


\title{The Hidden Topology of Network Entanglement}

\author{Debasish Sarker}
\email{dsarker@channing.harvard.edu}
\affiliation{%
Channing Division of Network Medicine, Brigham and Women’s Hospital,
Harvard Medical School, Boston, MA 02115, USA%
}


\begin{abstract}
Topology and geometry are inseparable in physical networks. Here we show that, after factoring out the topology-dependent number of eligible edge pairs, exchangeability makes the mean crossing propensity of those pairs exactly graph independent while preserving topology in fluctuations. We derive a microscopic, parameter-free theory for the response when topology and geometry are coupled. Synthetic and real three-dimensional networks confirm that structure hidden from the normalized mean survives in correlations that determine response.
\end{abstract}

\flushbottom
\maketitle


\paragraph{Introduction.---}

Physical networks must realize their connectivity in three-dimensional space, where the spatial embedding of nodes and links constrains the structures available to the network~\cite{BARTHELEMY20111, dehmamy2018structural, blagojevic2024three, PhysRevLett.133.077401}. In neuronal, vascular, root, fungal, and mitochondrial networks~\cite{stiso2018spatial, tekin2016vascular, fitter1991architectural, bebber2007biological, rafelski2013mitochondrial}, topology and geometry are therefore inseparable aspects of physical organization: connectivity determines which elements are connected, while spatial embedding determines how those connections are realized in space. These constraints can generate entangled configurations that restrict geometric rearrangements and influence physical response~\cite{liu2021isotopy, panagiotou2019topological}. Recent work introduced the average crossing number as a measure of entanglement in physical networks and showed that it depends on both network topology and spatial layout~\cite{PhysRevLett.133.077401}. Link length controls the likelihood of crossings, while network density, degree heterogeneity, and community structure modify the crossing opportunities available to the network.

Separating these topological and geometric contributions statistically remains difficult because changing the connectivity also changes which spatial positions are joined. Existing approaches have characterized their interplay through spatial constraints and wiring costs~\cite{BARTHELEMY20111}, isotopy and linking of physical-network embeddings~\cite{liu2021isotopy}, and projection-averaged crossing statistics~\cite{PhysRevLett.133.077401}. A complementary line of work has studied edge crossings under randomized vertex arrangements, deriving their expectation and variance in random linear arrangements and relating these statistics to graph structure~\cite{alemany2020edge}. These results characterize fluctuations of the crossing number itself, but do not address how crossing fluctuations couple to geometric observables or determine response when topology and geometry become statistically associated. This leaves a fundamental question: if a fixed topology and a fixed geometry are statistically decoupled, what information about the graph survives after the combinatorial number of crossing opportunities is factored out? If graph structure disappears from the mean crossing propensity of those opportunities, is that information truly lost, or does it remain hidden in the fluctuations?


\begin{figure}[!htb]
    \centering
    \includegraphics[width=1\linewidth]{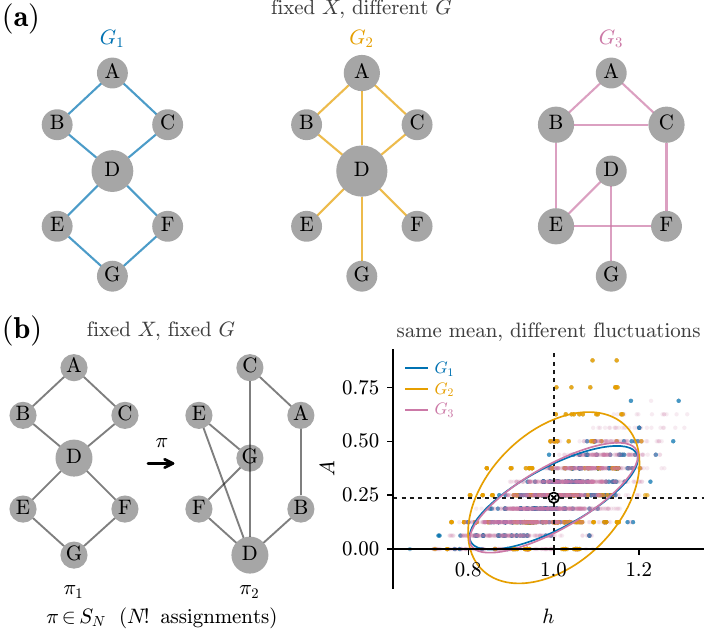}
    \caption{\textbf{Topology under spatial randomization.}
    (a) Three graph topologies, $G_1$, $G_2$, and $G_3$, on the same point cloud $X$.
    (b) For fixed $G$ and $X$, all $N!$ assignments $\pi\in S_N$ of vertices to positions are considered (left).
    The resulting ensembles have the common mean
    $(\langle h\rangle_0,\langle A\rangle_0)=(1,p_X)$
    but different fluctuations (right); ellipses show the covariance structure of each ensemble.}
    \label{fig:spatial_null}
\end{figure}

Here we show that exchangeability produces a sharp separation between topological opportunity, mean crossing propensity, and response. We fix both the graph and the spatial point cloud and uniformly randomize only the association between vertices and positions. Topology determines how many nonadjacent edge pairs are eligible to cross. Once this combinatorial factor is removed, however, the mean crossing propensity per eligible pair is exactly independent of graph topology: every eligible pair samples the same distribution of four-point geometries. Yet topology is not erased. It survives in correlations among overlapping graph objects and therefore in the fluctuations about the common mean. We resolve these fluctuations into geometric covariance classes weighted by graph-specific overlap multiplicities and obtain a parameter-free susceptibility that predicts how the normalized mean responds when topology and geometry are coupled. Topology determines how often the relevant microscopic correlations occur, while geometry determines their strength.

We test this prediction in controlled synthetic networks and in 15 three-dimensional physical networks spanning neuronal, vascular, mitochondrial, root, anthill, and fruit-fly brain systems~\cite{blagojevic2024three}. Across these structurally distinct networks, the microscopic theory quantitatively predicts the susceptibility without fitting the response. Independent sampling of the coupled ensemble then confirms that the susceptibility of the exchangeable state predicts the linear response away from the reference point. The framework therefore provides an exchangeable reference for assessing crossing-based entanglement relative to randomized topology--geometry association.


\paragraph{Theoretical framework.---}

We consider a simple graph $G=(V,E)$ with $N$ vertices and $L$ edges, and a fixed set of spatial positions
$X=\{\mathbf{x}_1,\ldots,\mathbf{x}_N\}\subset\mathbb{R}^3$.
An embedding is specified by a permutation $\pi\in S_N$ that assigns vertex $i$ to position $\mathbf{x}_{\pi(i)}$ [Fig.~\ref{fig:spatial_null}(a)]. Varying $\pi$ preserves both the graph and the point cloud and changes only their association.

Let $\mathcal{Q}(G)$ denote the set of $m$ pairs of nonadjacent edges. For each $q\in\mathcal{Q}(G)$, let $p_q(\pi)$ be the probability that the corresponding pair of segments crosses under an isotropically random projection~\cite{PhysRevLett.133.077401}. The projection-averaged crossing count is $C(\pi)=\sum_{q\in\mathcal{Q}(G)}p_q(\pi)$. Because the number $m$ of eligible pairs is itself topology dependent, we define the normalized crossing-based entanglement
\begin{equation}
    A(\pi)
    =
    \frac{C(\pi)}{m}
    =
    \frac{1}{m}
    \sum_{q\in\mathcal{Q}(G)}
    p_q(\pi).
    \label{eq:entanglement}
\end{equation}
Thus topology determines the number of crossing opportunities through $m$, while $A$ measures their mean crossing propensity, separating it from the combinatorial number of eligible edge pairs.

We first consider the exchangeable ensemble in which all $N!$ vertex-position assignments are equally likely, $P_0(\pi)=1/N!$. Every eligible edge pair contains four distinct vertices, so a uniform permutation maps its four labeled endpoints uniformly onto an ordered 4-tuple of distinct points of $X$. It follows that $\langle p_q\rangle_0=p_X$ for every $q$, where $p_X$ is the mean projection-averaged crossing probability over ordered 4-tuples of distinct points, with $x_1,x_2$ assigned to the endpoints of one edge and $x_3,x_4$ to those of the other. Hence
\begin{equation}
    \langle A\rangle_0=p_X .
    \label{eq:null_mean}
\end{equation}
Equation~(\ref{eq:null_mean}) shows that the mean crossing propensity per eligible pair is independent of graph topology under exchangeable vertex-position assignments. The corresponding unnormalized mean remains $\langle C\rangle_0=m p_X$ and retains the topology-dependent number of crossing opportunities. Different graphs on the same point cloud nevertheless have the same $\langle A\rangle_0$, as illustrated in Fig.~\ref{fig:spatial_null}(b). Exchangeability fixes only the mean contribution of each eligible pair. Correlations between graph objects can still depend on how those objects overlap, so graph structure removed from this normalized first moment can remain in the fluctuations.


\begin{figure}[!htb]
    \centering
    \includegraphics[width=1\linewidth]{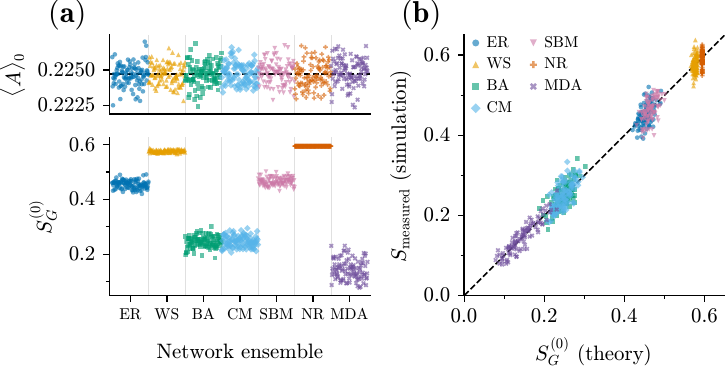}
    \caption{\textbf{Topology dependence of entanglement susceptibility.}
(a) Null-ensemble mean entanglement $\langle A\rangle_0$ (top) and susceptibility $S_G^{(0)}$ (bottom) for 100 network realizations from each of seven graph ensembles: Erd\H{o}s--R\'enyi (ER)~\cite{erdds1959random},
Watts--Strogatz (WS)~\cite{watts1998collective}, Barab\'asi--Albert (BA)~\cite{doi:10.1126/science.286.5439.509}, configuration model (CM)~\cite{BOLLOBAS1980311}, stochastic block model (SBM)~\cite{HOLLAND1983109}, near-regular (NR), and mediation-driven attachment (MDA)~\cite{SARKER2020109591}.
The dashed line indicates the mean $\langle A\rangle_0$; scatter about this mean reflects finite-sample estimation, whereas Eq.~(\ref{eq:null_mean}) is exact.
(b) The independently measured response $S_{\mathrm{measured}}$ is plotted against $S_G^{(0)}$.
The dashed line denotes $S_{\mathrm{measured}}=S_G^{(0)}$.}
    \label{fig:susceptibility}
\end{figure}


\paragraph{Topology--geometry coupling.---}

To probe this remaining topological information, we consider the total Euclidean edge length
$H_G(\pi)=\sum_{(i,j)\in E}|\mathbf{x}_{\pi(i)}-\mathbf{x}_{\pi(j)}|$
and define the dimensionless mean edge length
$h(\pi)=H_G(\pi)/(Ld_0)$, where
$d_0=\binom{N}{2}^{-1}\sum_{a<b}|\mathbf{x}_a-\mathbf{x}_b|$
is the mean pairwise distance of the point cloud. Under the exchangeable ensemble, every graph edge samples a uniformly chosen pair of distinct points, giving $\langle h\rangle_0=1$. Thus all graphs on the same cloud share
$(\langle h\rangle_0,\langle A\rangle_0)=(1,p_X)$, while their joint fluctuations can differ [Fig.~\ref{fig:spatial_null}(b)].

We couple topology to geometry by biasing the vertex-position assignments according to
\begin{equation}
    P_\kappa(\pi\mid G,X)
    =
    \frac{1}{Z_\kappa}
    \exp[-\kappa Lh(\pi)] .
    \label{eq:biased_ensemble}
\end{equation}
At $\kappa=0$, the exchangeable ensemble is recovered. Positive $\kappa$ favors assignments with shorter graph edges, while negative $\kappa$ favors longer ones. Neither the graph nor the point cloud is changed; $\kappa$ changes only the statistical weight of different vertex-position assignments.

For any observable $O$ with no explicit $\kappa$ dependence,
differentiating Eq.~(\ref{eq:biased_ensemble}) gives the
fluctuation--response relation~\cite{marconi2008fluctuation},
\[
\frac{d\langle O\rangle_\kappa}{d\kappa}
=
-L\,\operatorname{Cov}_\kappa(O,h).
\]
Applying this relation to $A$ and $h$ yields
\[
    \frac{d\langle A\rangle_\kappa}
         {d\langle h\rangle_\kappa}
    =
    \frac{\operatorname{Cov}_\kappa(A,h)}
         {\operatorname{Var}_\kappa(h)}.
\]
The response at $\kappa=0$ is therefore determined by fluctuations of the exchangeable ensemble. We denote this null susceptibility by
$S_G^{(0)}=\left.d\langle A\rangle_\kappa/d\langle h\rangle_\kappa\right|_{\kappa=0}$, with the dependence on the fixed point cloud $X$ understood. Unlike the common normalized mean $\langle A\rangle_0$, $S_G^{(0)}$ varies with the graph even at fixed $X$ [Fig.~\ref{fig:susceptibility}(a)]. To identify the origin of this graph dependence, we resolve the null fluctuations microscopically.


\paragraph{Topological susceptibility.---}

Writing
$A=m^{-1}\sum_q p_q$ and
$h=(Ld_0)^{-1}\sum_e\ell_e$, where $\ell_e$ is the Euclidean length of edge $e$, gives
\[
    \operatorname{Cov}_0(A,h)
    =
    \frac{1}{mLd_0}
    \sum_{q,e}
    \operatorname{Cov}_0(p_q,\ell_e).
\]
Each term couples an eligible edge pair $q$ to an edge $e$. Under exchangeability, its value is determined by the point cloud $X$ and by how $e$ overlaps the two edges forming $q$, but not by the identities of the vertices involved. Similarly, the terms in $\operatorname{Var}_0(h)$ are determined by $X$ and by whether two graph edges coincide, share a vertex, or are disjoint. The covariance sums can therefore be organized into a finite set of overlap classes.

This classification separates the geometric and topological contributions. Within each overlap class, exchangeability gives a common geometric covariance determined by $X$, while the graph enters through the multiplicity of that class. We denote the three crossing--length covariance classes that remain after the exact reduction by $C_s$, $C_\times$, and $C_1$, with corresponding graph-dependent multiplicities $\alpha_s$, $\alpha_\times$, and $\alpha_1$. For the edge-length fluctuations, $V_\ell$ denotes the variance of a single edge length and $C_{\rm adj}$ the covariance between two edge lengths sharing a vertex, with multiplicities $\beta_V$ and $\beta_{\rm adj}$, respectively. Grouping the covariance sums by these overlap classes and carrying out the exact reduction of the disconnected contributions gives
\begin{equation}
\begin{aligned}
    S_G^{(0)}
    &\equiv
    \left.
    \frac{d\langle A\rangle_\kappa}
         {d\langle h\rangle_\kappa}
    \right|_{\kappa=0}
    =
    \frac{\operatorname{Cov}_0(A,h)}
         {\operatorname{Var}_0(h)}
    \\[2pt]
    &=
    \frac{Ld_0}{m}
    \frac{
        \alpha_s C_s
        +\alpha_\times C_\times
        +\alpha_1 C_1
    }{
        \beta_V V_\ell
        +\beta_{\rm adj}C_{\rm adj}
    }.
\end{aligned}
\label{eq:susceptibility}
\end{equation}
The explicit definitions of the overlap classes and their multiplicities, together with the complete covariance reduction and exact finite-system verification, are given in the Supplemental Material (SM) (see SM sections 3--4).

Equation~(\ref{eq:susceptibility}) shows where the graph dependence removed from the normalized mean in Eq.~(\ref{eq:null_mean}) survives. Exchangeability fixes the geometric statistics within each overlap class, but not the number of times each class occurs in the graph. After the overall number of eligible pairs has been factored out, the normalized first moment loses its dependence on $G$, while the correlations governing the susceptibility retain graph-specific information through overlap multiplicities. When $\kappa$ couples topology to geometry, these null fluctuations determine the change in the normalized mean. Thus Eq.~(\ref{eq:susceptibility}) predicts the response of the coupled ensemble from the uncoupled ensemble without fitting a response coefficient.


\begin{figure}[!htb]
    \centering
    \includegraphics[width=1\linewidth]{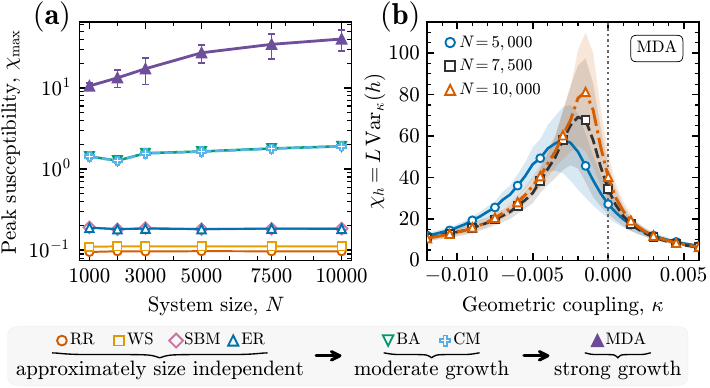}
    \caption{\textbf{Finite-size geometric susceptibility.}
    \textbf{(a)} Peak susceptibility $\chi_{\max}$ versus system size $N$
    for seven network families, evaluated on a common coupling grid for
    cross-architecture comparison.
    Points denote means over 20 independent graph realizations and error bars
    denote 95\% confidence intervals.
    \textbf{(b)} Geometric susceptibility $\chi_h(\kappa,N)$ for MDA networks
    at $N=5000$, $7500$, and $10000$, obtained from a refined coupling scan
    around the susceptibility peak to resolve its finite-size structure.
    Curves denote ensemble means over 20 graph realizations and shaded regions denote 95\% confidence intervals.}
    \label{fig:critical}
\end{figure}

Equation~(\ref{eq:susceptibility}) shows that fluctuations of the
exchangeable state determine its linear response. This raises a natural
finite-size question: do these fluctuations remain bounded as the system
grows, or can network architecture amplify them collectively? Applying the
same fluctuation--response relation to $h$ defines the geometric
susceptibility
\begin{equation}
    \chi_h(\kappa,N)
    \equiv
    L\,\operatorname{Var}_{\kappa}(h)
    =
    -\frac{d\langle h\rangle_\kappa}{d\kappa}.
    \label{eq:geometric_susceptibility}
\end{equation}
For each system size $N$, we characterize the maximal response by
$\chi_{\max}(N)\equiv\max_{\kappa}\chi_h(\kappa,N)$.
Its finite-size behavior reveals a pronounced dependence on network
architecture [Fig.~\ref{fig:critical}(a)]. Here RR denotes the random-regular control~\cite{2014_Random_Regular},
distinct from the near-regular (NR) ensemble used in
Fig.~\ref{fig:susceptibility}.
RR, WS, SBM, and ER remain approximately size independent, while BA and its
degree-preserving configuration-model counterpart CM show nearly identical
moderate growth. MDA exhibits much stronger amplification. The close
correspondence between BA and CM points away from growth history alone and
toward degree heterogeneity as the relevant structural ingredient. Consistent
with this interpretation, $\chi_{\max}$ across MDA realizations is strongly
associated with the degree second moment $\langle k^2\rangle$ ($r=0.997$),
with the association persisting within fixed system size.

The MDA susceptibility also develops a characteristic finite-size structure
[Fig.~\ref{fig:critical}(b)]. As $N$ increases, the peak grows and narrows,
while its location
$\kappa_{\rm peak}(N)\equiv\arg\max_{\kappa}\chi_h(\kappa,N)$
moves toward $\kappa=0$. This simultaneous growth, sharpening, and approach
to the exchangeable state constitutes a finite-size signature of
critical-like response, although it does not by itself establish a
thermodynamic phase transition. Thus topology that is absent from the
normalized exchangeable mean can re-emerge in response and become
collectively amplified in structurally heterogeneous networks (see SM section 6).



\begin{figure}[!htb]
    \centering
    \includegraphics[width=0.6\linewidth]{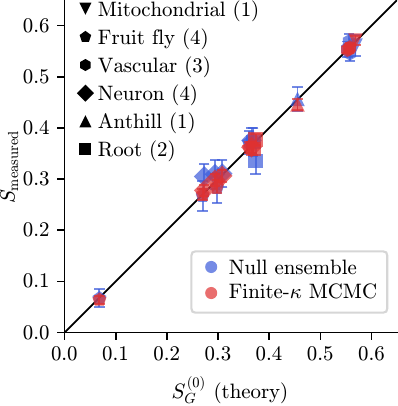}
    \caption{\textbf{Microscopic-theory prediction across real spatial networks.}
    Comparison of the microscopic-theory prediction with direct null-ensemble susceptibility measurements and with independently sampled finite-$\kappa$ MCMC response measurements.
    The null comparison tests the microscopic reduction within the exchangeable ensemble, whereas the finite-$\kappa$ comparison tests the predicted response using independent coupled-ensemble sampling.
    Numbers in parentheses indicate the number of distinct empirical networks in each network type, not the number of samples used to estimate an individual susceptibility.
    Error bars denote 95\% confidence intervals for the measured susceptibility of each individual network; details of the null-ensemble and finite-$\kappa$ analyses, including uncertainty estimation, are given in SM sections 8--9.
    The solid line denotes $S_{\mathrm{measured}}=S_G^{(0)}(\mathrm{theory})$.}
    \label{fig:empirical}
\end{figure}

\paragraph{Predictive validation.---}

We first test Eq.~(\ref{eq:susceptibility}) under controlled conditions using synthetic networks, varying topology while holding the spatial point cloud fixed. Across the seven graph ensembles in Fig.~\ref{fig:susceptibility}(a), the exchangeable mean $\langle A\rangle_0$ remains unchanged, whereas the susceptibility varies substantially with topology. The independently measured linear response follows the null susceptibility [Fig.~\ref{fig:susceptibility}(b)].

We next consider the 15 three-dimensional physical-network data sets
compiled and standardized by Blagojevi\'c and P\'osfai~\cite{blagojevic2024three} (see SM section 7). They fall into three broad topological classes: trees, comprising four individual neurons, two plant root systems, and an anthill imprint; lattice-like networks, comprising three vascular networks and a mitochondrial network; and linked trees, comprising four fruit-fly brain regions in which tree-like neuronal structures are connected by synapses.

For each network, we retain the graph and spatial point cloud and randomize only the vertex-position association. Figure~\ref{fig:empirical} compares the resulting null susceptibility with the microscopic prediction of Eq.~(\ref{eq:susceptibility}). Across all 15 networks, the microscopic prediction agrees closely with the directly measured null susceptibility, with $r=0.9935$ and a relative RMSE of $4.07\%$, without fitting a response coefficient; the prediction lies within the corresponding direct-null 95\% confidence interval for 13 of the 15 networks. Thus the microscopic decomposition captures the susceptibility across structurally distinct empirical graph--geometry pairs, while the controlled synthetic comparison at fixed $X$ isolates its graph dependence (see SM section 8).

We finally test whether a susceptibility determined entirely at the exchangeable point predicts the response of the coupled topology--geometry ensemble. We independently sample the biased ensemble in Eq.~(\ref{eq:biased_ensemble}) and measure the linear response of $\langle A\rangle$ to changes in $\langle h\rangle$ near $\kappa=0$. The measured ensemble response agrees quantitatively with the parameter-free susceptibility predicted from the uncoupled ensemble across all 15 empirical graph--geometry pairs, with $r=0.9993$ and a relative RMSE of $1.30\%$; all 15 predicted susceptibilities fall within the corresponding finite-coupling 95\% confidence intervals [Fig.~\ref{fig:empirical}; see SM section 9].


\paragraph{Discussion.---}

Our results establish a precise separation between topological opportunity,
mean crossing propensity, and response. In the exchangeable ensemble,
topology determines the number $m$ of nonadjacent edge pairs eligible to
cross, giving $\langle C\rangle_0=m p_X$, while the normalized mean
$\langle A\rangle_0=p_X$ contains no further dependence on graph topology.
That information nevertheless survives in correlations among overlapping
graph objects. Exchangeability fixes the geometric covariance associated
with each overlap class, while the graph determines how frequently each
class occurs, so macroscopic response emerges from the combinatorial
organization of microscopic geometric correlations. The agreement across
synthetic ensembles and structurally distinct physical networks shows that
this mechanism, rather than a particular susceptibility, is common across
these systems: graph structure removed from the normalized first moment
remains encoded in the fluctuations that determine response.

More broadly, our results connect two established ideas in statistical
physics. Fluctuation--response theory relates spontaneous fluctuations of an
unperturbed system to its response under a perturbation~\cite{marconi2008fluctuation}.
In network science, statistical null ensembles provide reference states in
which selected structural constraints are preserved while other features are
randomized~\cite{cimini2019statistical}. Our construction brings these ideas
together for spatial networks, where connectivity and spatial embedding are
intrinsically intertwined~\cite{BARTHELEMY20111}: the reference ensemble
preserves both the graph and the point cloud while randomizing only their
association. The resulting response relation does more than connect
fluctuations to susceptibility: it identifies what structural information
those fluctuations retain after the normalized mean crossing propensity has
become graph independent. This suggests a broader strategy for systems with
coupled structural and spatial degrees of freedom: preserve the constituent
structures, randomize their association, and use fluctuations about the
resulting reference state to identify information that is invisible to an
appropriately normalized mean but revealed by response.

The present theory also defines the scope of this construction and the
questions it leaves open. We hold the graph and point cloud fixed,
characterize entanglement through projection-averaged pair crossings, and
introduce topology--geometry coupling through total Euclidean edge length.
The exchangeable ensemble is a statistical reference state rather than a
model of physical network rearrangement; individual randomized embeddings
need not satisfy the geometric or mechanical constraints of the underlying
physical system. Equation~(\ref{eq:susceptibility}) determines the leading
response about this reference state, but does not by itself determine
behavior far from it. Extending the framework to fluctuating connectivity or
geometry, alternative geometric and topological observables, and other
physically motivated couplings would test how broadly the separation between
mean behavior and hidden structural response persists. Beyond linear
response, the finite-size behavior in Fig.~\ref{fig:critical} suggests that
these microscopic correlations can be collectively amplified in an
architecture-dependent manner. Whether this behavior develops into a genuine
thermodynamic transition, and which structural properties govern its
asymptotic scaling, remain open questions. More generally, our results show
that normalization and averaging can remove a visible signature of structure
without erasing the information carried by its correlations: structure
hidden from a normalized mean can reappear through response.


\begin{acknowledgments}
The author acknowledges support from the National Institutes of Health under Award No.~R01HL171141.
\end{acknowledgments}


\bibliography{ref.bib}

\end{document}